\documentclass[epjST_arxiv,final]{svjour_arxiv}

\usepackage{amsmath,epsfig,alltt,subfigure,graphicx,color}
\usepackage[sort&compress]{natbib}

\def\nn{\nonumber}

\def\s{\textrm{s}}
\def\S{\textrm{S}}
\def\D{\textrm{D}}

\def\psiLStoD{|\psi^L_{\S\to \D}|}
\def\psiCStoD{|\psi^C_{\S\to \D}|}
\def\sinc{\textrm{sinc}}

\def\sumLm{ \sum_{m=1}^{\infty}e^{-m^2\pi^2Dt/B^2}L_m(\mu_1,\mu_2) }

\def\sumLmmu{
	\sum_{m=1}^{\infty}e^{-m^2\pi^2Dt/B^2}L_m(\mu,\mu) }

\def\Eb{\begin{eqnarray}}
	\def\Ee{\end{eqnarray}}
\def\Ebsub{\begin{subequations}\begin{eqnarray}}
		\def\Eesub{\end{eqnarray}\end{subequations}}

\begin{document}

\title{Dynamic coherent backscattering of ultrasound in three-dimensional strongly-scattering media}

\author{L. A. Cobus\inst{1} \and B. A. van Tiggelen\inst{2}\inst{3} \and A. Derode\inst{4} \and J. H. Page\inst{1}\fnmsep\thanks{\email{john.page@umanitoba.ca}}} 

\institute{Department of Physics and Astronomy, University of Manitoba, Winnipeg, Manitoba R3T 2N2, Canada \and Universit\'{e} Grenoble Alpes, LPMMC, F-38000 Grenoble, France \and CNRS, LPMMC, F-38000 Grenoble, France \and Institut Langevin, ESPCI ParisTech, CNRS UMR 7587, Universit\'{e} Denis Diderot - Paris 7, 1 rue Jussieu, 75005 Paris, France}

\abstract{We present measurements of the diffusion coefficient of ultrasound in strongly scattering three-dimensional (3D) disordered media using the dynamic coherent backscattering (CBS) effect. Our experiments measure the CBS of ultrasonic waves using a transducer array placed in the far-field of a 3D slab sample of brazed aluminum beads surrounded by vacuum. We extend to 3D media the general microscopic theory of CBS that was developed initially for acoustic waves in 2D.  This theory is valid in the strong scattering, but still diffuse, regime that is realized in our sample, and is evaluated in the diffuse far field limit encountered in our experiments. By comparing our theory with the experimental data, we obtain an accurate measurement of the diffusion coefficient of ultrasound in our sample. We find that the value of $D$ is quite small, $0.74 \pm 0.03$ mm$^2/\mu$s, and comment on the implications of this slow transport for the energy velocity.} 

\maketitle

\section{Introduction}
Coherent backscattering (CBS) has for several decades been used to measure transport parameters of disordered media in the diffuse regime. The CBS effect is caused by interference between multiply-scattered waves travelling reciprocal paths inside a disordered medium \cite{Sheng2006}. Experimentally, this phenomenon may be observed as an enhancement (of around 2) in intensity at exact backscattering. Away from exact backscattering the CBS intensity profile decreases, forming a `cone' shape which contains valuable information about scattering parameters of the medium \cite{VanAlbada1985,Wolf1985,Kuga1985,Akkermans1988}. As has been observed experimentally for various types of diffuse waves and scattering media \cite{VanAlbada1985,Wolf1985,Kuga1985,Akkermans1988,Bayer1993,Tourin1997,Jonckheere2000,Wolf1988}, the width of the static (single frequency or time-integrated dynamic) CBS profile is directly related to the transport mean free path, $\l^*$. The dynamic (time-dependent) CBS profile provides opportunities to measure additional quantities. In the diffuse regime, the dynamic CBS profile can directly yield a measurement of the Boltzmann diffusion coefficient $D_B$ without the influence of absorption. Most measurements of $D_B$ using dynamic CBS have been performed for acoustic waves in 2D media \cite{Tourin1997,Mamou2005,Aubry2007,Bayer1993}.
However, the first acoustic study of CBS by Bayer et al. in 1993 \cite{Bayer1993} also investigated 3D samples, in which dynamic CBS from a very thick (effectively semi-infinite) gravel medium was observed. These data were interpreted using theory taken directly from electromagnetics, which includes assumptions which may not be justified for acoustics. Here, we present a microscopic derivation of dynamic CBS for acoustic waves in 3D, in which we also take into account the conditions encountered in experiments with ultrasonic transducer arrays. The theory takes into account the diffuse near field, which would be measured if a detector were placed at a distance from the sample that is comparable to, or less than, the width of the diffuse halo at the surface. In our experiments, the transducer array is placed much farther from the sample surface than this distance, so that the theory is specifically evaluated in the diffuse far field limit that applies to our data. We compare our theory with experimental measurements of time-dependent acoustic CBS from a 3D medium, and show that our approach yields an accurate measurement of the Boltzmann diffusion coefficient.

\section{Experiment}

\begin{figure}[t]
	\begin{subfigure}{}
		\centering
		\includegraphics[width=0.8\columnwidth]{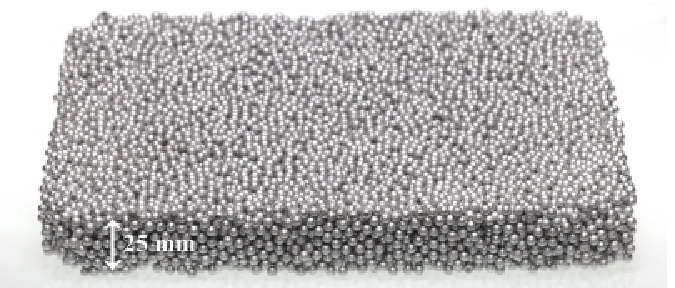}
	\end{subfigure}
	\begin{subfigure}{}
		\centering
		\includegraphics[width=0.8\columnwidth]{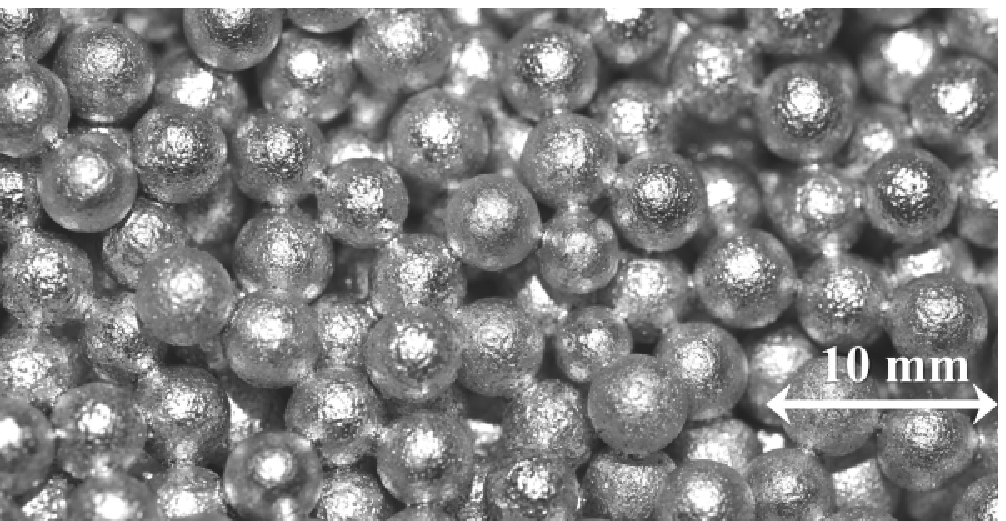}
		\caption{(a)  Sample L1 has a thickness of $L=25 \pm 2$ mm and a cross-section of 230$\times$250 mm$^2$.  (b)  The bead structure of sample L1.}
		\label{figa}
	\end{subfigure}
\end{figure}

Backscattered ultrasound was measured from a slab-shaped mesoglass sample composed of aluminum beads brazed together to form a disordered elastic network (Fig.~\ref{figa}). The bead volume fraction in the sample was $\sim 55$\%, and the mean bead diameter was 3.93 mm with a polydispersity of about 20\%, which helps to randomize bead positions. The sample has a cross-section of 230$\times$250 mm$^2$ much larger than its thickness $L= 25 \pm 2$ mm, which helps to minimize contributions from the edges of the sample when performing backscattering experiments. Other details of the sample characteristics have been described in Refs. \cite{Aubry2014, Cobus2016,Cobus2016phd}.

\begin{figure}[t]\centering
	\includegraphics[width=0.9\columnwidth]{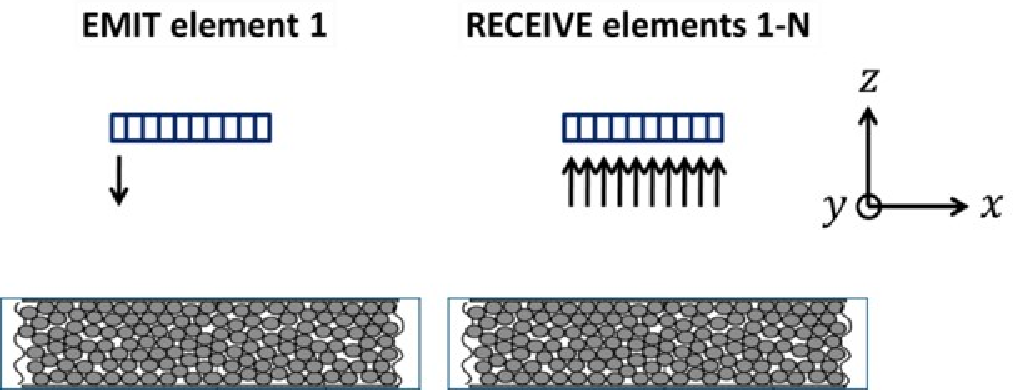}
	\caption{First step of the acquisition sequence for an ultrasonic array of $N$ elements.}
	\label{fig0}
\end{figure}

An ultrasonic array with central frequency $f_c = 1.6$ MHz was used to measure backscattered field from the sample. The experiment was done in a large plexiglass water tank, with sample and array immersed in water, parallel to each other and separated by a distance of $a=182$ mm. Before the experiment was performed, the sample was waterproofed, and for the duration of the experiment the pores between the beads were held under vacuum so that the propagation of ultrasonic waves inside the sample occurs only in the elastic network. Note that in this set-up, compressional and shear waves propagate in the sample, but at the transducers all excitations have been mode-converted back to compressional waves. As a result, the CBS is essentially a scalar phenomenon, despite the vector nature of the waves in the sample. The experimental acquisition process is sketched in Fig.~\ref{fig0}: a single element emits a short pulse, and then all (64) elements record the time-dependent backscattered field. By repeating this process, emitting with each element in turn, the time-dependent 'response matrix' was acquired \cite{Aubry2014, Tourin1997}. Configurational averaging was performed by translating the array parallel to the sample surface and acquiring the response matrices for 302 different positions. Prior to each experiment, careful checks were carried out to minimize and hopefully eliminate any spurious reflections that could have contributed to the backscattered field from the sample, including the small possible contribution from signals that had travelled through the sample, reflected off either a sample support or a tank wall, and then travelled back through the sample en route to the detector.

To study a particular frequency range, the data were filtered using a Gaussian envelope of standard deviation 0.025 MHz, centered in this case around $f=1.65$ MHz. This frequency has been shown to exhibit conventional diffusive behaviour of ultrasound \cite{Cobus2016phd} (as opposed to subdiffusive or localized behaviour, which has been studied for this sample at other frequencies \cite{Aubry2014, Cobus2016}). However, the scattering is still very strong at 1.65 MHz, as is evidenced by the significant contribution to the total backscattered intensity from recurrent scattering processes, which reduce the CBS enhancement below 2 \cite{Aubry2014}. On average, over all times investigated here (between $\sim 20 - 220$ $\mu$s), recurrent scattering constitutes as much as $37\%$ of the total backscattered intensity, and for the latest times (between $170 - 220$ $\mu$s), the observed recurrent scattering contribution is still more than $15\%$ \cite{Aubry2014,Cobus2016phd}. This contribution complicates the analysis of CBS, since it adds to the flat, angle-independent background intensity level (the intensity contribution given by Equation~\ref{fourDI} in the next section) \cite{Wiersma1995,vanTiggelen1995,Aubry2014}. The recurrent scattering contribution was removed from the total backscattered intensity using the approach developed by Aubry et al. \cite{Aubry2014}. The result of experiments and data-processing is a large set of configurationally-averaged, time-dependent backscattered intensity profiles $I(\rho,t)$, where $\rho$ is the distance between source and receiver elements of the ultrasonic array, and $t$ is time. To eliminate the effect of absorption, $I(\rho,t)$ was normalized by $I(0,t)$, since at time $t$ the effect of absorption is the same for both numerator and denominator of this ratio, and therefore should cancel \cite{Hu2008,Cobus2016}.

\section{Theory}
\begin{figure}[]\centering
	\includegraphics[width=0.8\columnwidth]{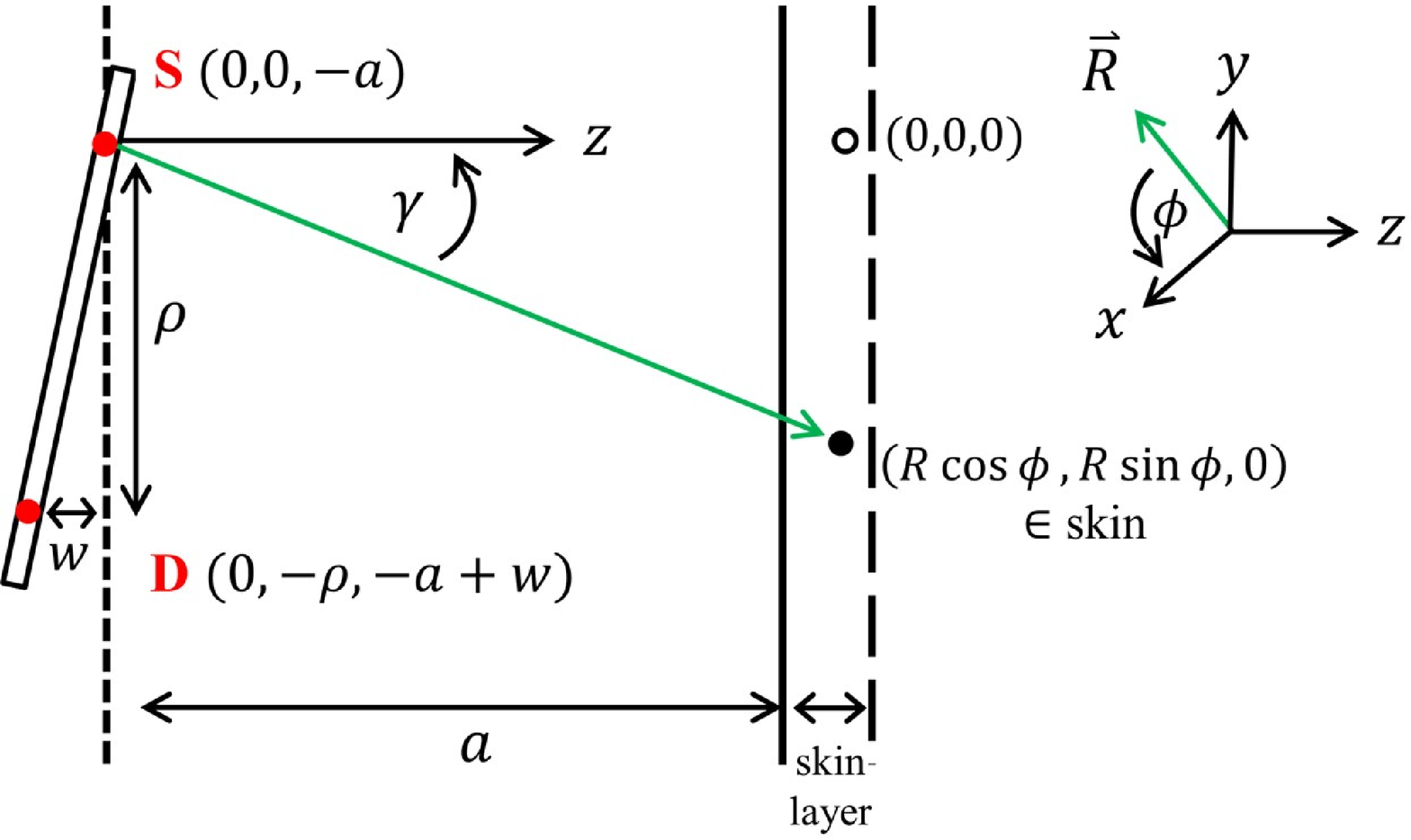}
	\caption{Schematic for the experimental geometry and coordinate systems. The source/detector plane (an ultrasonic array) is approximately parallel to sample surface, although source S and detector D may be separated in the $z$ direction by a small distance $w$ (exaggerated in the diagram for clarity).  Inside the sample, scattering is described by the coordinate system shown on the lower right. The origin of this $R,\phi$ system, $(0,0,0)$, is on the ray normal to the emitting transducer array element at S. Vector $\vec{R}$ is in the $xy$ plane, which is parallel to the input/output sample surface.}
	\label{fig1}
\end{figure}

Here we outline our theory for diffuse, strong scattering of acoustic waves in 3D samples. The geometry of the system is shown in Fig.~\ref{fig1}. Source S is positioned at $(0,0,-a)$, and detector D at $(0,-\rho,-a+w)$, where $a$ is the perpendicular distance between S and the sample surface, and $w$ describes any additional distance between S and D, in the $z$ direction. In an experiment with an ultrasonic array, $w$ is minimized as much as possible by aligning the array as parallel as possible to the sample surface, but it is still useful to be able to account for any residual misalignment in the theoretical analysis. The first scattering event occurs in the `skin layer' of the sample, at a distance equal on average to the scattering mean free path $\ell_s$. The theory presented here assumes that the scattering length is small, i.e. $a>>\ell_s$, and neglects any phase shift which might occur in the skin layer. Additionally, we assume that the problem is symmetric in $\phi$, and that $k_0a>>1$, where $k_0$ is the wave vector in the water in front of the sample.

We start with the general expression for intensity from a scattering medium:
\Eb
\langle|\psi_{\S\to \D}|^2\rangle &\propto& \int G(\vec{r}_1) G^*(\vec{r}_3)\Gamma(\vec{r}_1,\vec{r}_2,\vec{r}_3,\vec{r}_4)	
G(\vec{r}_2)G^*(\vec{r}_4)d\vec{r}_1 d\vec{r}_2 d\vec{r}_3 d\vec{r}_4,
\label{eq1}
\Ee
where propagation between source and sample is given by ensemble averaged Green's functions $G(\vec{r}_1)$ and $G(\vec{r}_3)$, and propagation between sample and detector is given by $G(\vec{r}_2)$ and $G(\vec{r}_4)$. In the far-field approximation $|\vec{R}-\vec{r}_1|\rightarrow\infty$, and in the diffusive regime where $k\ell_s>>1$ ($k=2\pi f/v_p$ is the wavevector in the medium, $v_p$ is the velocity of the longitudinal waves), the ensemble averaged Green's function between source and sample in 3D may be approximated as
\Eb
G(S\to{(\vec{R}_{1,3},z)}\in\textrm{skin})=-\frac{e^{i k\sqrt{a^2+R_{1,3}^2}} e^{-\frac{z}{2\mu\ell_s}}}{4\pi \sqrt{a^2+R_{1,3}^2}}
\label{GS},
\Ee
with $\mu_{1,3}=\cos\gamma=a/\sqrt{a^2+R_{1,3}^2}$. To more accurately express the intensity drop due to propagation from the last point in the sample at $(\vec{R}_{2,4},z)$, to detector D at $(\rho,-a+w)$, the ensemble averaged Green's function between sample and detector is written as
\Eb
\nonumber
G({(\vec{R}_{2,4},z)}\in\textrm{skin}\to D)=-e^{-\frac{z}{2\mu\ell_s}}\\
\times
\frac{e^{i k\sqrt{a^2+R_{2,4}^2}}}{4\pi \sqrt{(a-w)^2+R_{2,4}^2+\rho^2-2\rho R_{2,4}\cos\phi}}.
\label{GD}
\Ee
All scattering inside the medium is described by vertex $\Gamma(\vec{r}_1,\vec{r}_2,\vec{r}_3,\vec{r}_4)$. For the incoherent contribution to backscattered intensity, $\Gamma=F_{3D}(\vec{r}_1,\vec{r}_2)\delta_{1,3}\delta_{2,4}$, where in the ladder approximation (the diffuse regime), the 3D ladder kernel $F_{3D}$ is a solution to the diffusion equation. 
For a slab-shaped medium with partial reflection of energy at the boundaries, i.e. $F_{3D}=0$ at $z=z_0$, $z=L+z_0$, the 3D ladder kernel is \cite{C&J1995}
\Eb
&F_{3D}&(R, z_1, z_2, t) = C\frac{e^{-R^2/4D_B t}}{D_B t}
\sum_{m=1}^\infty e^{-m^2\pi^2D_B t/B^2}	\nn\\
&\times& \left[\cos \left(\frac{\pi m(z_1-z_2)}{B}\right) -\cos\left(\frac{\pi m(z_1+z_2-2z_0)}{B}\right) \right], \nn\\
& &
\Ee
where $D_B = v_e l^*/3$ is the Boltzmann diffusion coefficient in 3D (c.f. Ref.~\cite{Schriemer1997}) and $R=\left|\vec{R_1}-\vec{R_2}\right|$. The extrapolation length in 3D is $z_0=(2/3)(1+R_\textrm{refl})/(1-R_\textrm{refl})\ell^*$, effective sample thickness is defined as $B\equiv L+2z_0$, and $C$ is a constant. Through the boundary conditions, the transport mean free path $\ell^*$ has now been introduced, and is allowed to differ from the scattering mean free path $\ell_s$ (anisotropic scattering).

We now evaluate equation~\ref{eq1} for our particular experimental geometry. The integrations over $z$ are the same as in the theory for 2D \cite{Tourin1999} with some differences in the denominators, and once evaluated, give for the incoherent background intensity
\Eb
&&\psiLStoD^2(t) = \int_0^\infty dR_1 R_1\int_0^\infty dR_2 R_2 
\int_0^{2\pi}d\phi_1\int_0^{2\pi}d\phi_2 \nn\\
& \times & \frac{ \exp\left( -\frac{ R_1^2 + R_2^2 - 2R_1R_2cos\phi_{12}}{4Dt}\right) }{ \left(a^2 + R_1^2\right)(a^2 + R_2^2 + \rho^2 - 2\rho R_2\cos\phi_2) }
\sumLm,
\label{fourDI}
\Ee
where the $L_m$ terms are \cite{Tourin1999}
\Eb
L_m(\mu_{1}, \mu_2)=\frac{2(A_1+B_{1})(A_2 + B_2)}{(a_{1}^2 + b_m^2)(a_2^2 + b_m^2)}
\label{Lm}
\Ee
with $a_\textrm{1,2} = 1/l_s\mu_\textrm{1,2}$, $b_m = \pi m/B$,
\Eb
A_{1,2}(m) & = & \sin(b_m z_0)a_{1,2}\left[1 + \left(-a_{1,2}\right)^me^{-a_{1,2}L}\right]\nn\\
B_{1,2}(m) & = & \cos(b_m z_0)b_m\left[1 + (-a_{1,2})^me^{-a_{1,2}L}\right]\nn.
\Ee

To calculate the backscattered intensity due to the CBS effect (the {\it coherent} contribution), the scattering vertex is $\Gamma=F_{3D}(\vec{r}_1,\vec{r}_2)\delta_{2,3}\delta_{1,4}$. This means that contributions from interferences between pairs of reciprocal paths through the sample are taken into account. The expression for the coherent contribution to backscattered intensity thus includes an extra phase factor compared to the incoherent case: 
\Eb
&\psiCStoD&^2(t) = \int_0^\infty dR_1 R_1\int_0^\infty dR_2 R_2 
\int_0^{2\pi}d\phi_1\int_0^{2\pi}d\phi_2\nn\\
& \times & \cos\left[k_0(\mu_1-\mu_2)w + k_0(s_1\sin\phi_1 - s_2\sin\phi_2)\rho\right]\nn\\
& \times & \frac{ \exp\left( -\frac{ R_1^2 + R_2^2 - 2R_1R_2\cos\phi_{12}}{4Dt}\right) }{ (a^2 + R_1^2)(a^2 + R_2^2 + \rho^2 - 2\rho R_2\cos\phi_2) } \sumLm,
\label{fourDC}
\Ee
for $w<<\sqrt{a^2+R_{1,2}^2}$ and $\rho<<\sqrt{a^2+R_{1,2}^2}$, and where $s_{1,2}=\sin{\gamma}=\sqrt{1-{\mu_{1,2}}^2}$. To partially evaluate the integral, we apply the \textit{diffuse far-field assumption} $a^2>>4D_B t$. This corresponds physically to a diffuse halo on the sample that is much smaller than the distance between the sample and transducer. For strong scattering, it is important to assess whether or not the experimental data obey this assumption. In our experiment, the longest times collected are around 220 $\mu$s, and for a diffusion coefficient $D_B$ of order  $0.7$ mm$^2/\mu$s (see the next section), the assumption $a^2=(182$ mm$)^2\approx 33000$ mm$^2>>4D_B t\approx4(0.7$ mm$^2/\mu$s$)($220 $\mu$s$)\approx 600$ mm$^2$ holds. In our experiment, the array misalignment $w$ was too small to be measurable, and thus is set to zero from now on. Because of the exponential factor in the integral, the dominant contributions in the diffuse far field approximation come from the points $\vec{R_1}$ and $\vec{R_2}= \vec{R_1} + \Delta \vec{R}$  separated by small distances relative to $a$, so that it is convenient to change coordinates from $\vec{R_1}$ and $\vec{R_2}$  to $\vec{R}$ and $\Delta\vec{R}$.  Then, Eq. 7 can be simplified by expanding the argument of the cosine factor to first order in $\Delta\vec{R}$ and integrating over $\Delta\vec{R}$. The coherent backscattered intensity then simplifies to
\Eb
&\psiCStoD^2&(t)  \cong \int_0^\infty dR R \int_0^{2\pi}d\phi 
\frac{ \exp\left(\mu^2 \frac{(k_0\rho)^2Dt}{a^2}\left[s^4\sin^2\phi + 1 - 2s^2\sin^2\phi \right]\right) }
{ (a^2 + R^2)(a^2 + R^2 + \rho^2 - 2\rho R\cos\phi) }\nn\\
& \times &
\sumLmmu,
\label{twoDCnear}
\Ee
where $\mu=a/\sqrt{a^2+R^2}$~\cite{wzero}. In the \textit{diffuse coda}, $t > \tau_{D} = B^2/\pi^2D_B$ ($\tau_D$ is the diffusion time), only the $m=1$ term in Eq.~\ref{Lm} survives, simplifying the calculation. This $m=1$ term can be further simplified if the \textit{optically thick slab} approximation, $B>>\ell^*$, applies. In our slab, the smallest possible value of $B=L+2z_0=L+2(2/3)l^*(1+R_\textrm{refl})/(1-R_\textrm{refl})$ is $B=25 \textrm{ mm}+2(2/3)(4 \textrm{ mm})(1+0.65)/(1-0.65)\approx50$ mm. Thus, the approximation of $B>>\ell^*$ is obeyed since $B=50\textrm{ mm}>>4\textrm{ mm}$. However, in our experimental situation we measure up to 220 $\mu$s, so most times considered are smaller than the smallest possible $\tau_{D} \approx (50 \textrm{ mm})^2/\pi^2(0.7 \textrm{ mm}^2/\mu\textrm{s}) \approx$ 360 $\mu\s$. Thus, the diffuse coda approximation does not hold, and all of the terms in the $L_m$ series of Eq.~\ref{Lm} are included in our calculations.

As shown by Eq.~\ref{twoDCnear}, the shape of the CBS dynamic cone is determined by the dimensionless parameter $k_0\rho(D_B t/a^2)$. At each time $t$ the CBS intensity profile has an almost Gaussian shape, with a width $(k\rho)_\textrm{FWHM}$ that depends on time as
\Eb
(k_0\rho)_\textrm{FWHM}= \Gamma a^2/D_B t,
\label{k0rho}
\Ee
where $\Gamma$ is a dimensionless constant. This coherent contribution adds to the uniform incoherent background given by Eq.~\ref{fourDI}. At $\rho=0$, the coherent and incoherent contributions to total backscattered intensity should be equal.

For the sake of completeness, we note that the static CBS intensity profile may be found (in the absence of absorption) simply by integrating Eqs.~\ref{fourDI} and~\ref{twoDCnear} over all time. Assuming $B>>\ell^*$, the ratio of coherent to incoherent intensity can be expressed analytically as
\Eb
\frac{\psiCStoD^2}{\psiLStoD^2} & \cong & \frac{a}{\frac{2}{3}\ell_s + z_0}
\int_0^1 \mu d\mu \int_0^{2\pi} \frac{d\phi}{2\pi} \frac{1 - \exp\left(-Q\frac{2\mu \ell_s + 2z_0}{a}\right)}{Q},
\label{twoDstat}
\Ee
where
\Eb
Q = \mu\sqrt{\mu^4\sin^2\phi + \cos^2\phi}\ k_0\rho. \nn
\Ee
As has been found for other systems, the width of the static profile depends inversely on $\ell^*$ \cite{vanderMark1988,Akkermans1988,Tourin1997}. \\

For technical reasons, experimental results for the static CBS intensity profile are not included in this work. This is mainly due to the fact that at early times the signal is dominated by large specular reflections which could not be completely eliminated with the recurrent scattering filter \cite{Aubry2014}. In addition, we do not have data at sufficiently late times to accurately calculate the static cone; due to the strong scattering nature of the sample, the dynamic backscattered `cones' do not narrow very quickly, but the range of accessible times in the measurements is limited to times before the arrival of the next echo between sample surface and transducer (after a time interval of $2a/v_\textrm{water}$ $\approx 240$ $ \mu$s). These considerations demonstrate the advantages of our dynamic CBS measurements, which do not require data for all times but can yield accurate results as long as the {\it range} of times experimentally available is sufficient to demonstrate the dynamics of the CBS profiles.

\section{Extensions of theory to account for experimental conditions}
Several modifications to the above theory were made for a more accurate comparison between theory and experiment. The theory so far assumes point sources and detectors, whereas experimentally the source and detector have a finite rectangular shape, with width $W=0.25$ mm (in the $x$ direction of Fig. \ref{fig1}) much less than height $H=12$ mm ($y$ direction of Fig. \ref{fig1}). This means that the directivity (directional dependence) of each element should be taken into account, especially in the $x$ direction where the spreading of waves due to diffraction from the narrow elements is not insignificant. Here we estimate the directivity along $x$ using the ideal profile for a rectangular transducer of width $W$ 
in the far field:
\Eb
\psi(\theta) = \psi(0)\sinc\left(\frac{\pi W\sin\theta}{\lambda}\right)
\Ee
The same expression can be used in the $y$ direction, so that for an array in 3D, the total angular sensitivity is
\Eb
\Omega(\theta_x,\theta_y)=\sinc^2\left(\frac{\pi W\sin\theta_x}{\lambda}\right)\sinc^2\left(\frac{\pi H\sin\theta_x}{\lambda}\right)
\Ee
The correction is incorporated in Eq.~\ref{twoDCnear} as:
\Eb
&\psiCStoD&^2(t) = \int_0^\infty dR R \int_0^{2\pi}d\phi\ 
\Omega(\theta_{x,S},\theta_y)\Omega(\theta_{x,D},\theta_y)\nn\\
& \times &
\frac{ \exp\left(\mu_1^2 \frac{(k\rho)^2Dt}{a^2}\left[s^4\sin^2\phi + 1 - 2s^2\sin^2\phi \right]\right) }
{ (a^2 + R^2)(a^2 + R^2 + \rho^2 - 2\rho R\cos\phi) }\nn\\
& \times &
\sumLm,
\label{twoDCnear_dircorr}
\Ee
where $\theta_{x,S}$, $\theta_{x,D}$ and $\theta_y$ are found from
\begin{displaymath}\begin{array}{ccccc}
\tan(\theta_{x,S}) & = & \frac{R\cos\phi}{a}\nn\\
\tan(\theta_{x,D}) & = & \frac{R\cos\phi - \xi}{a}\nn\\
\tan(\theta_y)     & = & \frac{R\sin\phi}{a}. \nn
\end{array}\end{displaymath}
An additional correction was performed to account for the height of the array elements, since the height of $H=12$ mm means that signal is being collected over a significantly greater area than was supposed by our theory. The effect is not so large that interference cancellation of ultrasonic field at the array surface is important, so an integration over detected intensity is sufficient to account for the influence of element height. The correction consists of performing an explicit (numerical) integration of the intensity distribution $|\psi(\rho,t)|^2$ over all possible source points ($y_1$) and receiver points ($y_2$), by calculating an effective $\rho$ for each pair of points:
\Eb
|\psi(\rho,t)|^2_\textrm{corr} &=& \int_{-H/2}^{H/2}dy_1\int_{-H/2}^{H/2}dy_2 |\psi(\sqrt{\rho^2+(y_2-y_1)},t)|^2.
\Ee
The same procedure (with the replacement of $H$ by $W$) is performed to account for the finite width of the array elements. Overall, the geometrical corrections presented in this section do not change the global trend of Eq.~\ref{k0rho}, but do change the multiplicative factor $\Gamma$.

\section{Fitting and Results}

\begin{figure*}[]\centering
	\includegraphics[width=\textwidth]{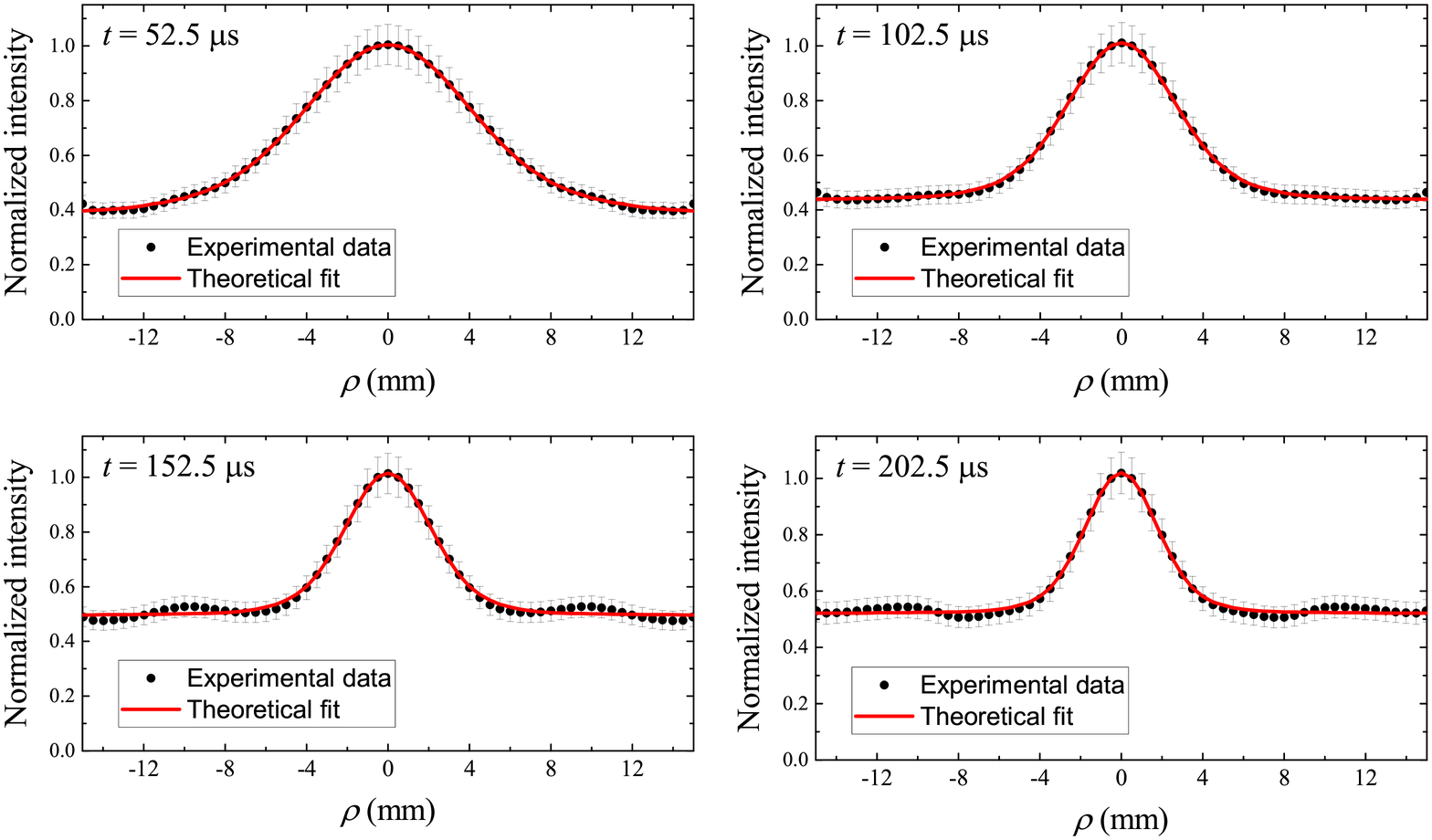}
	\caption{Experimental CBS profiles (symbols) with fits from diffusion theory (lines) for four representative times: 52.5 $\mu s$, 102.5 $\mu s$, 152.5 $\mu s$, and 202.5 $\mu s$. Error bars represent the experimental uncertainty in the configurational average.}
	\label{fig2}
\end{figure*}

Representative experimental CBS profiles are shown in Fig.~\ref{fig2} (symbols). In principle, after the removal of the recurrent scattering contribution, the incoherent background intensity level should be at 0.5. Our experimental data deviate from this value slightly at some times, with the deviations being especially small at late times. This may be caused by an inaccuracy in the recurrent scattering filter, especially at early times where the initial specular reflection is very large and difficult to remove entirely \cite{Aubry2014}. Additionally, our theory shows that the enhancement factor may be slightly changed due to the finite size of the array elements.

To measure the diffusion coefficient $D_B$, the experimental CBS profiles $I_{\textrm{exp}}(\rho,t_\textrm{exp})$ were fit with the predictions of the diffusion theory outlined in the previous section. The theory calculations require several parameters as input, including scattering mean free path $\ell_s$ and reflection coefficient $R_\textrm{refl}$. From measurements of the coherent ballistic pulse in transmission \cite{Page1997}, we can determine $\ell_s \simeq 1.1$ mm. We also measure the longitudinal phase velocity $v_L\simeq2.8$ mm$/\mu$s \cite{Cobus2016phd} inside the sample, which is required to calculate $R_\textrm{refl}$. This calculation, based on methods developed by \cite{Zhu1991,Page1995,Ryzhik1996,Turner1995}, assumes that after a few scattering events, there is equipartition of energy between all polarizations of waves inside the sample, allowing equipartition to be taken into account when determining $R_\textrm{refl}$. In this calculation, the phase velocity of the dominant shear (transverse) waves inside the sample, $v_T$, is estimated to be $v_T\sim v_L/2\approx 1.4$ mm/$\mu$s.  Because shear waves dominate inside the sample, but longitudinal waves are detected outside the sample, the average reflection coefficient is large, $R_\textrm{refl} \approx 0.75$.

\begin{figure}[]\centering
	\includegraphics[width=0.8\columnwidth]{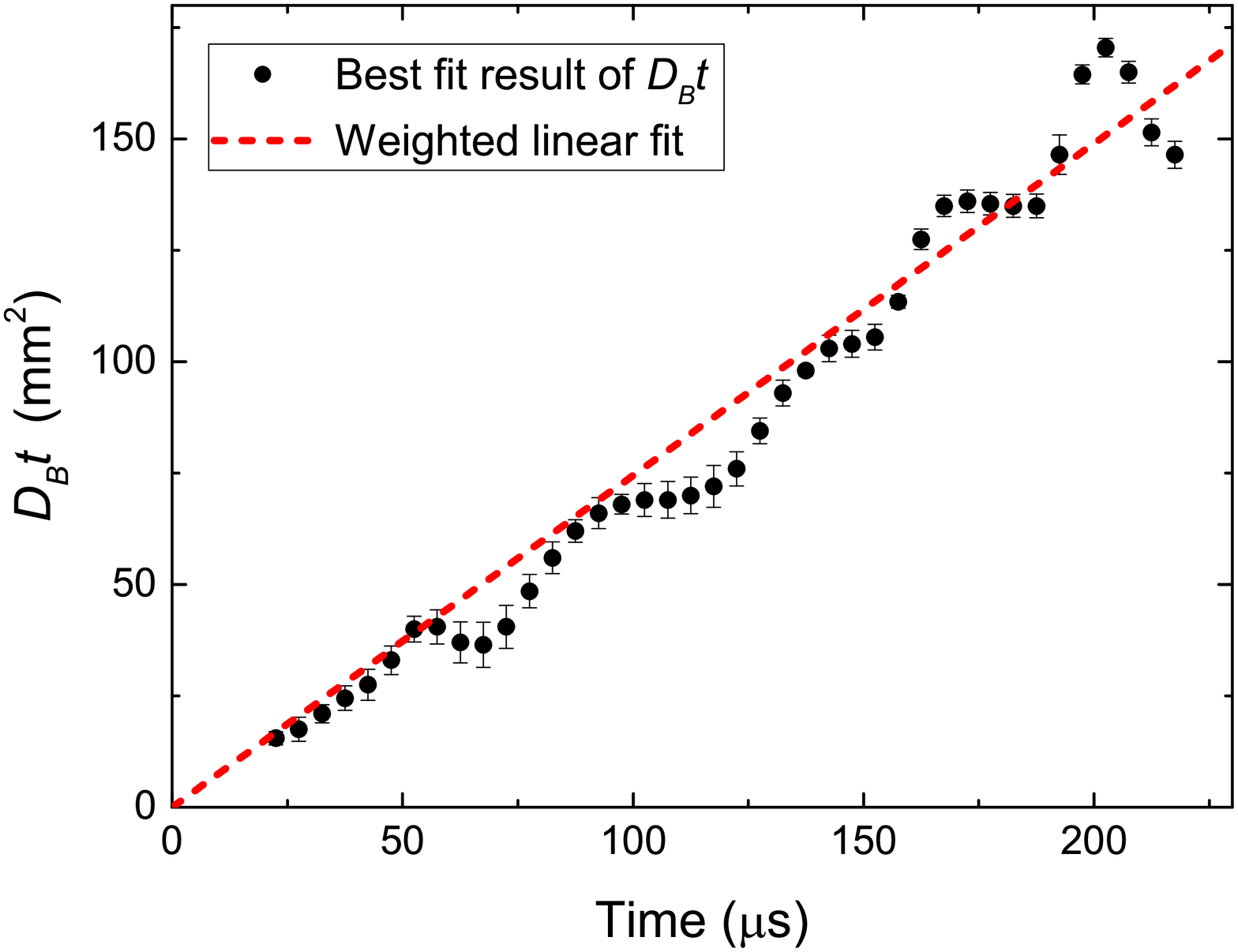}
	\caption{Results of the method used to determine the diffusion coefficient. Shown here are the $D_B t_\textrm{theory}$ values obtained from the best fit of the theory to the rho-dependent experimental CBS profiles at each time (symbols). The slope of a weighted linear fit to the data (dashed line) gives the overall value of $D_B$. Notable parameters used to calculate the theory in this plot are $\ell_s = 1.1$ mm, $\ell^* = 4$ mm, and $R_{\textrm{refl}}=0.75$ mm.  From this analysis of the data, we measure a value of $D_B=0.74\pm 0.03$ mm$^2/\mu$s. Error bars on each data point are determined from the goodness of fit of $I_{\textrm{theory}}(\rho,D_Bt)$ to $I_{\textrm{exp}}(\rho,t_\textrm{exp})$ (see text).}
	\label{fig3}
\end{figure}

Theoretical backscattering profiles $I_{\textrm{theory}}(\rho,D_Bt)$ were calculated as a function of parameter $D_Bt$ (diffusion coefficient multiplied by time) using Equations \ref{fourDI}, \ref{twoDCnear}, and \ref{twoDCnear_dircorr}. Then, all experimental CBS profiles were compared to all theoretical profiles, i.e. for each time $t_\textrm{exp}$, $I_{\textrm{exp}}(\rho,t_\textrm{exp})$ was fit with each theoretical CBS profile $I_{\textrm{theory}}(\rho,D_Bt)$. In this way, a best-fit value of $D_B t$ is determined for each $t_\textrm{exp}$. Figure~\ref{fig2} shows representative experimental CBS profiles for four different times $t_\textrm{exp}$, along with the best theoretical fits. Figure~\ref{fig3} shows the $D_B t$ values resulting from each best fit versus $t_\textrm{exp}$. The slope of a linear fit to these points gives a direct estimate of $D_B$. The data are well-described by the linear fit, confirming the prediction that $\Delta\rho^{-2}\propto D_B t$.

Since there is some uncertainty in the values of input parameters $R_{\textrm{refl}}$ and $\ell^*$ (c.f. \cite{Cobus2016,Cobus2016phd}), fitting was performed over the range of physically reasonable values for these parameters: $\ell^*=4-8$ mm and $R_{\textrm{refl}}=0.65-0.85$. It was found that over these ranges, the best-fit value of $D_B$ only varies by less than $0.01$ mm$^2/\mu$s, which is less than the fitting uncertainty in the determination of $D_B$ for any pair of these parameters ($\pm0.02$).  This shows that the measurement of $D_B$ by dynamic CBS is largely insensitive to the precise values of $\ell^*$ and $R_{\textrm{refl}}$, which only enter into the theory via the boundary conditions. Thus, our uncertainty in these parameters is unimportant for determining $D_B$ reliably, and the fitting of the CBS data with the theory presented here gives an accurate and precise measurement of $D_B=0.74\pm 0.03$ mm$^2/\mu$s. This value agrees with results from the measurement on the same sample of the transverse spread of the transmitted intensity, which gives $D_B = 0.71 \pm 0.02$ mm$^2/\mu$s \cite{Cobus2016phd}.  It is worth noting that the transmitted transverse width $=\sqrt{4Dt}$ is known to be independent of absorption and boundary conditions \cite{Page1995,Hu2008}, so that the excellent agreement between these reflection and transmission methods further confirms the accuracy of our present analysis of the dynamic CBS profiles.

The value of $D_B$ seems small if we compare the measured value with a very rough calculation using estimates of the equipartitioned velocity ($1.6$ mm$^2/\mu$s, which is close to the shear wave velocity) and transport mean free path (4 mm); this would give $D_B\sim 2$. Given the plausible range of $\ell^*$ values, which are supported by independent transmission experiments, we infer that the energy velocity $v_E$ itself must be very small; we find from our analysis that $v_E$ ($=3 D_B/\ell^*$) is between $0.2$ and $ 0.6$ mm/$\mu$s. Such values are much smaller (around $2.5-7$ times smaller) than either the shear or equipartitioned velocities of elastic waves in the sample (and $5-15$ times smaller than the longitudinal velocity), directly indicating how very slow the transport of energy by diffuse waves is in this strongly scattering sample.  This result is in striking contrast with the surprising large values of $v_E$ (approximately $3 - 5$ times \emph{larger} than the velocity of longitudinal waves) previously deduced from the analysis of data for a similar sample in the localization regime \cite{Hu2008}. This suggests that the large values of transport velocities observed previously are associated with Anderson localization.

Our results for $D_B$ demonstrate that careful consideration of sample and experimental details is necessary to obtain an accurate measurement of $D_B$. Without taking into account details about the geometry of the experiment, the theoretical prediction for the diffusion coefficient is given simply by $\Delta\rho^{-2} =k^2 D_B t$ (see the previous section, and \cite{Bayer1993,Tourin1997}). If this relation were used to measure $D_B$ from our experimental data, we would obtain $D_B = 0.5$ mm$^2\mu$s, which differs by 30\% from the actual value.

\section{Conclusions}

In this work, we have studied the dynamic coherent backscattering of ultrasound from a 3D medium in the diffusion regime. We used an ultrasonic transducer array to measure the backscattered ultrasound from a slab-shaped `mesoglass', enabling us to perform a very substantial amount of configurational averaging and to use a sophisticated filtering technique to remove, for times $>20$ $\mu$s, other contributions from specular reflections, single scattering and recurrent scattering.  As a result, accurate measurements of the dynamic CBS intensity profiles were obtained as a function of time and space/angle. We have described a microsopic derivation of the theory of CBS for acoustic waves in 3D, which provides an excellent description of the experimental data. The fitting of our CBS data with theory enables an absorption-free measurement of the diffusion coefficient of ultrasound in our sample, which was determined to be $D_B=0.74\pm 0.03$ mm$^2/\mu$s and which agrees with results from separate transmission experiments.

\acknowledgement
Roger Maynard was an inspiration in our community. We are all deeply grateful to Roger for so many pleasant moments and so many enlightening discussions, as well as for his enthusiastic support and encouragement over many years. In the context of this paper, J.H.P. would also like to thank Roger for introducing him to the first observations of coherent backscattering of acoustic waves in 1993, thereby initiating J.H.P.'s interest in studying this effect with ultrasound and motivating the current work. We would like to thank Matthew Hasselfield for his contributions to the code used to calculate the theoretical predictions for the CBS profiles, and for the guidance provided by his previous analysis of preliminary data on different samples. We would also like to thank Victor Mamou for carrying out similar experiments and first analyses on 2D synthetic samples. 
	
J.H.P. and L.A.C. acknowledge the support of NSERC (Discovery Grant RGPIN/9037-2001, Canada Government Scholarship, and Michael Smith Foreign Study Supplement), the Canada Foundation for Innovation and the Manitoba Research and Innovation Fund (CFI/MRIF, LOF Project 23523). A.D. benefited from funding by LABEXWIFI (Laboratory of Excellence ANR-10-LABX-24), within the French Program Investments for the Future under Reference No. ANR-10-IDEX-0001-02 PSL*. B.v.T. and J.H.P. are also grateful for support from the PICS program of the CNRS (project Ultra-ALT) and the Agence Nationale de la Recherche (grant ANR-14-CE26-0032 LOVE).
\endacknowledgement

\bibliography{DynamicCBS_DiffuseBvTtheory_2016}

\begin{thebibliography}{27}%
\makeatletter
\providecommand \@ifxundefined [1]{%
 \@ifx{#1\undefined}
}%
\providecommand \@ifnum [1]{%
 \ifnum #1\expandafter \@firstoftwo
 \else \expandafter \@secondoftwo
 \fi
}%
\providecommand \@ifx [1]{%
 \ifx #1\expandafter \@firstoftwo
 \else \expandafter \@secondoftwo
 \fi
}%
\providecommand \natexlab [1]{#1}%
\providecommand \enquote  [1]{``#1''}%
\providecommand \bibnamefont  [1]{#1}%
\providecommand \bibfnamefont [1]{#1}%
\providecommand \citenamefont [1]{#1}%
\providecommand \href@noop [0]{\@secondoftwo}%
\providecommand \href [0]{\begingroup \@sanitize@url \@href}%
\providecommand \@href[1]{\@@startlink{#1}\@@href}%
\providecommand \@@href[1]{\endgroup#1\@@endlink}%
\providecommand \@sanitize@url [0]{\catcode `\\12\catcode `\$12\catcode
  `\&12\catcode `\#12\catcode `\^12\catcode `\_12\catcode `\%12\relax}%
\providecommand \@@startlink[1]{}%
\providecommand \@@endlink[0]{}%
\providecommand \url  [0]{\begingroup\@sanitize@url \@url }%
\providecommand \@url [1]{\endgroup\@href {#1}{\urlprefix }}%
\providecommand \urlprefix  [0]{URL }%
\providecommand \Eprint [0]{\href }%
\providecommand \doibase [0]{http://dx.doi.org/}%
\providecommand \selectlanguage [0]{\@gobble}%
\providecommand \bibinfo  [0]{\@secondoftwo}%
\providecommand \bibfield  [0]{\@secondoftwo}%
\providecommand \translation [1]{[#1]}%
\providecommand \BibitemOpen [0]{}%
\providecommand \bibitemStop [0]{}%
\providecommand \bibitemNoStop [0]{.\EOS\space}%
\providecommand \EOS [0]{\spacefactor3000\relax}%
\providecommand \BibitemShut  [1]{\csname bibitem#1\endcsname}%
\let\auto@bib@innerbib\@empty
\bibitem [{\citenamefont {Sheng}(2006)}]{Sheng2006}%
  \BibitemOpen
  \bibfield  {author} {\bibinfo {author} {\bibfnamefont {P.}~\bibnamefont
  {Sheng}},\ }\href@noop {} {\emph {\bibinfo {title} {{Introduction to Wave
  Scattering, Localization and Mesoscopic Phenomena}}}},\ \bibinfo {edition}
  {2nd}\ ed.,\ edited by\ \bibinfo {editor} {\bibfnamefont {R.}~\bibnamefont
  {Hull}}, \bibinfo {editor} {\bibfnamefont {R.~M.~J.}\ \bibnamefont {Osgood}},
  \bibinfo {editor} {\bibfnamefont {J.}~\bibnamefont {Parisi}}, \ and\ \bibinfo
  {editor} {\bibfnamefont {H.}~\bibnamefont {Warlimont}}\ (\bibinfo
  {publisher} {Springer},\ \bibinfo {address} {Berlin},\ \bibinfo {year}
  {2006})\BibitemShut {NoStop}%
\bibitem [{\citenamefont {van Albada}\ and\ \citenamefont
  {Lagendijk}(1985)}]{VanAlbada1985}%
  \BibitemOpen
  \bibfield  {author} {\bibinfo {author} {\bibfnamefont {M.~P.}\ \bibnamefont
  {van Albada}}\ and\ \bibinfo {author} {\bibfnamefont {A.}~\bibnamefont
  {Lagendijk}},\ }\href@noop {} {\bibfield  {journal} {\bibinfo  {journal}
  {Physical Review Letters}\ }\textbf {\bibinfo {volume} {55}},\ \bibinfo
  {pages} {2692} (\bibinfo {year} {1985})}\BibitemShut {NoStop}%
\bibitem [{\citenamefont {Wolf}\ and\ \citenamefont {Maret}(1985)}]{Wolf1985}%
  \BibitemOpen
  \bibfield  {author} {\bibinfo {author} {\bibfnamefont {P.~E.}\ \bibnamefont
  {Wolf}}\ and\ \bibinfo {author} {\bibfnamefont {G.}~\bibnamefont {Maret}},\
  }\href@noop {} {\bibfield  {journal} {\bibinfo  {journal} {Physical Review
  Letters}\ }\textbf {\bibinfo {volume} {55}},\ \bibinfo {pages} {2696}
  (\bibinfo {year} {1985})}\BibitemShut {NoStop}%
\bibitem [{\citenamefont {Kuga}\ \emph {et~al.}(1985)\citenamefont {Kuga},
  \citenamefont {Tsang},\ and\ \citenamefont {Ishimaru}}]{Kuga1985}%
  \BibitemOpen
  \bibfield  {author} {\bibinfo {author} {\bibfnamefont {Y.}~\bibnamefont
  {Kuga}}, \bibinfo {author} {\bibfnamefont {L.}~\bibnamefont {Tsang}}, \ and\
  \bibinfo {author} {\bibfnamefont {A.}~\bibnamefont {Ishimaru}},\ }\href@noop
  {} {\bibfield  {journal} {\bibinfo  {journal} {Journals of the Optical
  Society of America A Communications}\ }\textbf {\bibinfo {volume} {2}},\
  \bibinfo {pages} {3} (\bibinfo {year} {1985})}\BibitemShut {NoStop}%
\bibitem [{\citenamefont {Akkermans}\ \emph {et~al.}(1988)\citenamefont
  {Akkermans}, \citenamefont {Wolf}, \citenamefont {Maynard},\ and\
  \citenamefont {Maret}}]{Akkermans1988}%
  \BibitemOpen
  \bibfield  {author} {\bibinfo {author} {\bibfnamefont {E.}~\bibnamefont
  {Akkermans}}, \bibinfo {author} {\bibfnamefont {P.~E.}\ \bibnamefont {Wolf}},
  \bibinfo {author} {\bibfnamefont {R.}~\bibnamefont {Maynard}}, \ and\
  \bibinfo {author} {\bibfnamefont {G.}~\bibnamefont {Maret}},\ }\href@noop {}
  {\bibfield  {journal} {\bibinfo  {journal} {J. Phys. France}\ }\textbf
  {\bibinfo {volume} {49}},\ \bibinfo {pages} {77} (\bibinfo {year}
  {1988})}\BibitemShut {NoStop}%
\bibitem [{\citenamefont {Bayer}\ and\ \citenamefont
  {Niederdr\"{a}nk}(1993)}]{Bayer1993}%
  \BibitemOpen
  \bibfield  {author} {\bibinfo {author} {\bibfnamefont {G.}~\bibnamefont
  {Bayer}}\ and\ \bibinfo {author} {\bibfnamefont {T.}~\bibnamefont
  {Niederdr\"{a}nk}},\ }\href@noop {} {\bibfield  {journal} {\bibinfo
  {journal} {Phys. Rev. Lett.}\ }\textbf {\bibinfo {volume} {70}},\ \bibinfo
  {pages} {3884} (\bibinfo {year} {1993})}\BibitemShut {NoStop}%
\bibitem [{\citenamefont {Tourin}\ \emph {et~al.}(1997)\citenamefont {Tourin},
  \citenamefont {Derode}, \citenamefont {Roux}, \citenamefont {van Tiggelen},\
  and\ \citenamefont {Fink}}]{Tourin1997}%
  \BibitemOpen
  \bibfield  {author} {\bibinfo {author} {\bibfnamefont {A.}~\bibnamefont
  {Tourin}}, \bibinfo {author} {\bibfnamefont {A.}~\bibnamefont {Derode}},
  \bibinfo {author} {\bibfnamefont {P.}~\bibnamefont {Roux}}, \bibinfo {author}
  {\bibfnamefont {B.~A.}\ \bibnamefont {van Tiggelen}}, \ and\ \bibinfo
  {author} {\bibfnamefont {M.}~\bibnamefont {Fink}},\ }\href@noop {} {\bibfield
   {journal} {\bibinfo  {journal} {Phys. Rev. Lett.}\ }\textbf {\bibinfo
  {volume} {79}},\ \bibinfo {pages} {3637} (\bibinfo {year}
  {1997})}\BibitemShut {NoStop}%
\bibitem [{\citenamefont {Jonckheere}\ \emph {et~al.}(2000)\citenamefont
  {Jonckheere}, \citenamefont {Muller}, \citenamefont {Kaiser}, \citenamefont
  {Miniatura},\ and\ \citenamefont {Delande}}]{Jonckheere2000}%
  \BibitemOpen
  \bibfield  {author} {\bibinfo {author} {\bibfnamefont {T.}~\bibnamefont
  {Jonckheere}}, \bibinfo {author} {\bibfnamefont {C.}~\bibnamefont {Muller}},
  \bibinfo {author} {\bibfnamefont {R.}~\bibnamefont {Kaiser}}, \bibinfo
  {author} {\bibfnamefont {C.}~\bibnamefont {Miniatura}}, \ and\ \bibinfo
  {author} {\bibfnamefont {D.}~\bibnamefont {Delande}},\ }\href@noop {}
  {\bibfield  {journal} {\bibinfo  {journal} {Physical Review Letters}\
  }\textbf {\bibinfo {volume} {85}},\ \bibinfo {pages} {4269} (\bibinfo {year}
  {2000})}\BibitemShut {NoStop}%
\bibitem [{\citenamefont {Wolf}\ \emph {et~al.}(1988)\citenamefont {Wolf},
  \citenamefont {Maret}, \citenamefont {Akkermans},\ and\ \citenamefont
  {Maynard}}]{Wolf1988}%
  \BibitemOpen
  \bibfield  {author} {\bibinfo {author} {\bibfnamefont {P.~E.}\ \bibnamefont
  {Wolf}}, \bibinfo {author} {\bibfnamefont {G.}~\bibnamefont {Maret}},
  \bibinfo {author} {\bibfnamefont {E.}~\bibnamefont {Akkermans}}, \ and\
  \bibinfo {author} {\bibfnamefont {R.}~\bibnamefont {Maynard}},\ }\href@noop
  {} {\bibfield  {journal} {\bibinfo  {journal} {J. Phys. France}\ }\textbf
  {\bibinfo {volume} {49}},\ \bibinfo {pages} {63} (\bibinfo {year}
  {1988})}\BibitemShut {NoStop}%
\bibitem [{\citenamefont {Mamou}(2005)}]{Mamou2005}%
  \BibitemOpen
  \bibfield  {author} {\bibinfo {author} {\bibfnamefont {V.}~\bibnamefont
  {Mamou}},\ }\emph {\bibinfo {title} {Caract\'{e}risation Ultrasonore
  D'\'{e}chantillons H\'{e}t\'{e}rog\`{e}nes Multiplement Diffuseurs}},\
  \href@noop {} {\bibinfo {type} {Doctoral thesis}},\ \bibinfo  {school}
  {Universit\'{e} Paris VII} (\bibinfo {year} {2005})\BibitemShut {NoStop}%
\bibitem [{\citenamefont {Aubry}\ and\ \citenamefont
  {Derode}(2007)}]{Aubry2007}%
  \BibitemOpen
  \bibfield  {author} {\bibinfo {author} {\bibfnamefont {A.}~\bibnamefont
  {Aubry}}\ and\ \bibinfo {author} {\bibfnamefont {A.}~\bibnamefont {Derode}},\
  }\href@noop {} {\bibfield  {journal} {\bibinfo  {journal} {Phys. Rev. E.}\
  }\textbf {\bibinfo {volume} {75}} (\bibinfo {year} {2007})}\BibitemShut
  {NoStop}%
\bibitem [{\citenamefont {Aubry}\ \emph {et~al.}(2014)\citenamefont {Aubry},
  \citenamefont {Cobus}, \citenamefont {Skipetrov}, \citenamefont {van
  Tiggelen}, \citenamefont {Derode},\ and\ \citenamefont {Page}}]{Aubry2014}%
  \BibitemOpen
  \bibfield  {author} {\bibinfo {author} {\bibfnamefont {A.}~\bibnamefont
  {Aubry}}, \bibinfo {author} {\bibfnamefont {L.~A.}\ \bibnamefont {Cobus}},
  \bibinfo {author} {\bibfnamefont {S.~E.}\ \bibnamefont {Skipetrov}}, \bibinfo
  {author} {\bibfnamefont {B.~A.}\ \bibnamefont {van Tiggelen}}, \bibinfo
  {author} {\bibfnamefont {A.}~\bibnamefont {Derode}}, \ and\ \bibinfo {author}
  {\bibfnamefont {J.~H.}\ \bibnamefont {Page}},\ }\href@noop {} {\bibfield
  {journal} {\bibinfo  {journal} {Phys. Rev. Lett.}\ }\textbf {\bibinfo
  {volume} {112}},\ \bibinfo {pages} {043903} (\bibinfo {year}
  {2014})}\BibitemShut {NoStop}%
\bibitem [{\citenamefont {Cobus}\ \emph {et~al.}(2016)\citenamefont {Cobus},
  \citenamefont {Aubry}, \citenamefont {Skipetrov}, \citenamefont {van
  Tiggelen}, \citenamefont {Derode},\ and\ \citenamefont {Page}}]{Cobus2016}%
  \BibitemOpen
  \bibfield  {author} {\bibinfo {author} {\bibfnamefont {L.~A.}\ \bibnamefont
  {Cobus}}, \bibinfo {author} {\bibfnamefont {A.}~\bibnamefont {Aubry}},
  \bibinfo {author} {\bibfnamefont {S.~E.}\ \bibnamefont {Skipetrov}}, \bibinfo
  {author} {\bibfnamefont {B.~A.}\ \bibnamefont {van Tiggelen}}, \bibinfo
  {author} {\bibfnamefont {A.}~\bibnamefont {Derode}}, \ and\ \bibinfo {author}
  {\bibfnamefont {J.~H.}\ \bibnamefont {Page}},\ }\href@noop {} {\bibfield
  {journal} {\bibinfo  {journal} {Phys. Rev. Lett.}\ }\textbf {\bibinfo
  {volume} {116}} (\bibinfo {year} {2016})}\BibitemShut {NoStop}%
\bibitem [{\citenamefont {Cobus}(2016)}]{Cobus2016phd}%
  \BibitemOpen
  \bibfield  {author} {\bibinfo {author} {\bibfnamefont {L.~A.}\ \bibnamefont
  {Cobus}},\ }\emph {\bibinfo {title} {Anderson Localization and Anomalous
  Transport of Ultrasound in Disordered Media}},\ \href@noop {} {\bibinfo
  {type} {Doctoral thesis}},\ \bibinfo  {school} {University of Manitoba}
  (\bibinfo {year} {2016})\BibitemShut {NoStop}%
\bibitem [{\citenamefont {Wiersma}\ \emph {et~al.}(1995)\citenamefont
  {Wiersma}, \citenamefont {van Albada}, \citenamefont {van Tiggelen},\ and\
  \citenamefont {Lagendijk}}]{Wiersma1995}%
  \BibitemOpen
  \bibfield  {author} {\bibinfo {author} {\bibfnamefont {D.~S.}\ \bibnamefont
  {Wiersma}}, \bibinfo {author} {\bibfnamefont {M.~P.}\ \bibnamefont {van
  Albada}}, \bibinfo {author} {\bibfnamefont {B.~A.}\ \bibnamefont {van
  Tiggelen}}, \ and\ \bibinfo {author} {\bibfnamefont {A.}~\bibnamefont
  {Lagendijk}},\ }\href@noop {} {\bibfield  {journal} {\bibinfo  {journal}
  {Phys. Rev. Lett.}\ }\textbf {\bibinfo {volume} {74}} (\bibinfo {year}
  {1995})}\BibitemShut {NoStop}%
\bibitem [{\citenamefont {van Tiggelen}\ \emph {et~al.}(1995)\citenamefont {van
  Tiggelen}, \citenamefont {Wiersma},\ and\ \citenamefont
  {Lagendijk}}]{vanTiggelen1995}%
  \BibitemOpen
  \bibfield  {author} {\bibinfo {author} {\bibfnamefont {B.~A.}\ \bibnamefont
  {van Tiggelen}}, \bibinfo {author} {\bibfnamefont {D.~A.}\ \bibnamefont
  {Wiersma}}, \ and\ \bibinfo {author} {\bibfnamefont {A.}~\bibnamefont
  {Lagendijk}},\ }\href@noop {} {\bibfield  {journal} {\bibinfo  {journal}
  {Europhys. Lett.}\ }\textbf {\bibinfo {volume} {30}} (\bibinfo {year}
  {1995})}\BibitemShut {NoStop}%
\bibitem [{\citenamefont {Hu}\ \emph {et~al.}(2008)\citenamefont {Hu},
  \citenamefont {Strybulevych}, \citenamefont {Page}, \citenamefont
  {Skipetrov},\ and\ \citenamefont {van Tiggelen}}]{Hu2008}%
  \BibitemOpen
  \bibfield  {author} {\bibinfo {author} {\bibfnamefont {H.}~\bibnamefont
  {Hu}}, \bibinfo {author} {\bibfnamefont {A.}~\bibnamefont {Strybulevych}},
  \bibinfo {author} {\bibfnamefont {J.~H.}\ \bibnamefont {Page}}, \bibinfo
  {author} {\bibfnamefont {S.~E.}\ \bibnamefont {Skipetrov}}, \ and\ \bibinfo
  {author} {\bibfnamefont {B.~A.}\ \bibnamefont {van Tiggelen}},\ }\href
  {\doibase 10.1038/nphys1101} {\bibfield  {journal} {\bibinfo  {journal} {Nat.
  Phys.}\ }\textbf {\bibinfo {volume} {4}},\ \bibinfo {pages} {945} (\bibinfo
  {year} {2008})}\BibitemShut {NoStop}%
\bibitem [{\citenamefont {Carslaw}\ and\ \citenamefont
  {Jaeger}(1995)}]{C&J1995}%
  \BibitemOpen
  \bibfield  {author} {\bibinfo {author} {\bibfnamefont {H.~S.}\ \bibnamefont
  {Carslaw}}\ and\ \bibinfo {author} {\bibfnamefont {J.~C.}\ \bibnamefont
  {Jaeger}},\ }\href@noop {} {\emph {\bibinfo {title} {Conduction of Heat in
  Solids}}},\ \bibinfo {edition} {2nd}\ ed.\ (\bibinfo  {publisher} {Oxford
  University Press},\ \bibinfo {year} {1995})\BibitemShut {NoStop}%
\bibitem [{\citenamefont {Schriemer}\ \emph {et~al.}(1997)\citenamefont
  {Schriemer}, \citenamefont {Cowan}, \citenamefont {Page}, \citenamefont
  {Sheng}, \citenamefont {Liu},\ and\ \citenamefont {Weitz}}]{Schriemer1997}%
  \BibitemOpen
  \bibfield  {author} {\bibinfo {author} {\bibfnamefont {H.~P.}\ \bibnamefont
  {Schriemer}}, \bibinfo {author} {\bibfnamefont {M.~L.}\ \bibnamefont
  {Cowan}}, \bibinfo {author} {\bibfnamefont {J.~H.}\ \bibnamefont {Page}},
  \bibinfo {author} {\bibfnamefont {P.}~\bibnamefont {Sheng}}, \bibinfo
  {author} {\bibfnamefont {Z.}~\bibnamefont {Liu}}, \ and\ \bibinfo {author}
  {\bibfnamefont {D.~A.}\ \bibnamefont {Weitz}},\ }\href@noop {} {\bibfield
  {journal} {\bibinfo  {journal} {Phys. Rev. Lett.}\ }\textbf {\bibinfo
  {volume} {79}} (\bibinfo {year} {1997})}\BibitemShut {NoStop}%
\bibitem [{\citenamefont {Tourin}(1999)}]{Tourin1999}%
  \BibitemOpen
  \bibfield  {author} {\bibinfo {author} {\bibfnamefont {A.}~\bibnamefont
  {Tourin}},\ }\emph {\bibinfo {title} {Diffusion multiple et renversement du
  temps des ondes ultrasonores}},\ \href@noop {} {\bibinfo {type} {Doctoral
  thesis}},\ \bibinfo  {school} {Universit\'{e} Paris VII} (\bibinfo {year}
  {1999})\BibitemShut {NoStop}%
\bibitem [{wze()}]{wzero}%
  \BibitemOpen
  \href@noop {} {}\bibinfo {note} {The details of this simplification may be
  found in Appendix 4A of~\cite{Cobus2016phd}}\BibitemShut {NoStop}%
\bibitem [{\citenamefont {van~der Mark}\ \emph {et~al.}(1988)\citenamefont
  {van~der Mark}, \citenamefont {van Albada},\ and\ \citenamefont
  {Lagendijk}}]{vanderMark1988}%
  \BibitemOpen
  \bibfield  {author} {\bibinfo {author} {\bibfnamefont {M.~B.}\ \bibnamefont
  {van~der Mark}}, \bibinfo {author} {\bibfnamefont {M.~P.}\ \bibnamefont {van
  Albada}}, \ and\ \bibinfo {author} {\bibfnamefont {A.}~\bibnamefont
  {Lagendijk}},\ }\href@noop {} {\bibfield  {journal} {\bibinfo  {journal}
  {Phys. Rev. B}\ }\textbf {\bibinfo {volume} {37}} (\bibinfo {year}
  {1988})}\BibitemShut {NoStop}%
\bibitem [{\citenamefont {Page}\ \emph {et~al.}(1997)\citenamefont {Page},
  \citenamefont {Schriemer}, \citenamefont {Jones}, \citenamefont {Sheng},\
  and\ \citenamefont {Weitz}}]{Page1997}%
  \BibitemOpen
  \bibfield  {author} {\bibinfo {author} {\bibfnamefont {J.~H.}\ \bibnamefont
  {Page}}, \bibinfo {author} {\bibfnamefont {H.~P.}\ \bibnamefont {Schriemer}},
  \bibinfo {author} {\bibfnamefont {I.~P.}\ \bibnamefont {Jones}}, \bibinfo
  {author} {\bibfnamefont {P.}~\bibnamefont {Sheng}}, \ and\ \bibinfo {author}
  {\bibfnamefont {D.~A.}\ \bibnamefont {Weitz}},\ }\href@noop {} {\bibfield
  {journal} {\bibinfo  {journal} {Phys. A.}\ }\textbf {\bibinfo {volume}
  {241}},\ \bibinfo {pages} {64} (\bibinfo {year} {1997})}\BibitemShut
  {NoStop}%
\bibitem [{\citenamefont {Zhu}\ \emph {et~al.}(1991)\citenamefont {Zhu},
  \citenamefont {Pine},\ and\ \citenamefont {Weitz}}]{Zhu1991}%
  \BibitemOpen
  \bibfield  {author} {\bibinfo {author} {\bibfnamefont {J.}~\bibnamefont
  {Zhu}}, \bibinfo {author} {\bibfnamefont {D.}~\bibnamefont {Pine}}, \ and\
  \bibinfo {author} {\bibfnamefont {D.~A.}\ \bibnamefont {Weitz}},\ }\href@noop
  {} {\bibfield  {journal} {\bibinfo  {journal} {Phys. Rev. A}\ }\textbf
  {\bibinfo {volume} {44}} (\bibinfo {year} {1991})}\BibitemShut {NoStop}%
\bibitem [{\citenamefont {Page}\ \emph {et~al.}(1995)\citenamefont {Page},
  \citenamefont {Schriemer}, \citenamefont {Bailey},\ and\ \citenamefont
  {Weitz}}]{Page1995}%
  \BibitemOpen
  \bibfield  {author} {\bibinfo {author} {\bibfnamefont {J.~H.}\ \bibnamefont
  {Page}}, \bibinfo {author} {\bibfnamefont {H.~P.}\ \bibnamefont {Schriemer}},
  \bibinfo {author} {\bibfnamefont {A.~E.}\ \bibnamefont {Bailey}}, \ and\
  \bibinfo {author} {\bibfnamefont {D.~A.}\ \bibnamefont {Weitz}},\ }\href@noop
  {} {\bibfield  {journal} {\bibinfo  {journal} {Phys. Rev. E}\ }\textbf
  {\bibinfo {volume} {52}},\ \bibinfo {pages} {3106} (\bibinfo {year}
  {1995})}\BibitemShut {NoStop}%
\bibitem [{\citenamefont {Ryzhik}\ \emph {et~al.}(1996)\citenamefont {Ryzhik},
  \citenamefont {Papanicolaou},\ and\ \citenamefont {Keller}}]{Ryzhik1996}%
  \BibitemOpen
  \bibfield  {author} {\bibinfo {author} {\bibfnamefont {L.}~\bibnamefont
  {Ryzhik}}, \bibinfo {author} {\bibfnamefont {G.}~\bibnamefont
  {Papanicolaou}}, \ and\ \bibinfo {author} {\bibfnamefont {J.~B.}\
  \bibnamefont {Keller}},\ }\href@noop {} {\bibfield  {journal} {\bibinfo
  {journal} {Wave Motion}\ }\textbf {\bibinfo {volume} {24}},\ \bibinfo {pages}
  {327} (\bibinfo {year} {1996})}\BibitemShut {NoStop}%
\bibitem [{\citenamefont {Turner}\ and\ \citenamefont
  {Weaver}(1995)}]{Turner1995}%
  \BibitemOpen
  \bibfield  {author} {\bibinfo {author} {\bibfnamefont {J.~A.}\ \bibnamefont
  {Turner}}\ and\ \bibinfo {author} {\bibfnamefont {R.~L.}\ \bibnamefont
  {Weaver}},\ }\href@noop {} {\bibfield  {journal} {\bibinfo  {journal} {J.
  Acoust. Soc. Am.}\ }\textbf {\bibinfo {volume} {98}},\ \bibinfo {pages}
  {2801} (\bibinfo {year} {1995})}\BibitemShut {NoStop}%
\end{thebibliography}%

\end{document}